# Generalized eigenvalue stabilization for immersed explicit dynamics

Tim Bürchner[*1], Lars Radtke[2], Sascha Eisenträger[3], Alexander Düster[4],
Ernst Rank[1,5], Stefan Kollmannsberger[6,1] and Philipp Kopp[6]

[1]Chair of Civil and Building Engineering, Technische Universität München
[2]Chair of Structural Mechanics, University of Rostock
[3]Institute of Materials, Technologies and Mechanics, Otto von Guericke University Magdeburg
[4]Institute for Ship Structural Design and Analysis, Technische Universität Hamburg
[5]Institute of Advanced Study, Technische Universität München
[6]Chair of Data Science in Civil Engineering, Bauhaus-Universität Weimar

## Abstract

Explicit time integration for immersed finite element discretizations severely suffers from the influence of poorly cut elements. In this contribution, we propose a generalized eigenvalue stabilization (GEVS) strategy for the element mass matrices of cut elements to cure their adverse impact on the critical time step size of the global system. We use spectral basis functions, specifically $C^0$ continuous Lagrangian interpolation polynomials defined on Gauss-Lobatto-Legendre (GLL) points, which, in combination with its associated GLL quadrature rule, yield high-order convergent diagonal mass matrices for uncut elements. Moreover, considering cut elements, we combine the proposed GEVS approach with the finite cell method (FCM) to guarantee definiteness of the system matrices. However, the proposed GEVS stabilization can directly be applied to other immersed boundary finite element methods. Numerical experiments demonstrate that the stabilization strategy achieves optimal convergence rates and recovers critical time step sizes of equivalent boundary-conforming discretizations. This also holds in the presence of weakly enforced Dirichlet boundary conditions using either Nitsche's method or penalty formulations.

## 1. Introduction

Immersed boundary finite element methods promise to pave the way to perform fast and accurate, but above all automated static and dynamic simulations involving complex geometries, without the need for boundary-conforming mesh generation, which typically involves more manual input. Most approaches follow a common philosophy: the physical domain of interest is embedded in a larger, geometrically simple domain that can be easily discretized, for instance, using a Cartesian grid. However, cut elements, i.e., elements intersected by the physical boundary, may introduce adverse effects. In particular elements with a very small support in the physical domain can lead to a system matrix with unfeasibly large condition numbers. In static simulations, the main concern is the resulting low convergence rate of iterative linear solvers, whereas in explicit dynamic simulations the critical time step size suffers [1]. Various stabilization approaches have been introduced to mitigate these issues. The finite cell method (FCM) uses material stabilization to add artificial mass and stiffness to the fictitious part of cut elements [2, 3]. Eigenvalue stabilization approaches shift the lower part of the eigenvalue spectrum to higher values using rank-one modifications [4–6]. CutFEM approaches constrain cut elements by weakly coupling their solution to full elements using ghost penalty formulations [7, 8]. Further prominent stabilization techniques include the aggregated finite element method (FEM) [9, 10], cgFEM [11], or the shifted boundary method [12, 13]. In the context of trimmed isogeometric analysis, extended B-splines [14, 15] and other polynomial extensions [16–18] have been proposed, as well as the isogeometric finite cell analysis (IGA-FCM) [19–21].

---
[*]`tim.buerchner@tum.de`, Corresponding author

In addition to the stabilization methods discussed above, specialized time integration schemes can be used to solve immersed hyperbolic problems efficiently. These schemes can overcome stability restrictions due to cut elements, which would otherwise severely limit the time step size. Immersed spectral discretizations yield a mass matrix with a special structure: a large diagonal part from the uncut elements, and a smaller consistent part from the cut elements. Tailored time integration schemes can exploit this structure. Examples include local time stepping [22, 23] and implicit-explicit (IMEX) time integration [24–27]. A comparative study in [28] investigates the computational performance of immersed spectral and isogeometric discretizations on a three-dimensional rotated cube example, highlighting the potential of combining advanced discretizations with tailored time integration strategies.

In the paper at hand, we address solving the immersed wave equation with explicit time integration, particularly with the central difference method (CDM). We employ high-order, $C^0$ continuous Lagrangian interpolation polynomials defined on Gauss-Lobatto-Legendre (GLL) points [29–31], as problems with smooth solutions benefit from the spectral convergence of polynomial approximations — i.e., the approximation error decreases exponentially with increasing polynomial degree. The spectral element method (SEM) [32, 33] combines this polynomial basis with the GLL quadrature rule, i.e., a nodal quadrature is utilized [34, 35]. This results in a diagonal mass matrix without a loss but even a gain in accuracy, see [36, 37]. Its immersed counterpart, the spectral cell method (SCM) [38, 3, 39], requires specialized integration schemes in the cut elements, leading to a consistent mass matrix in these elements. Material stabilization prevents the critical time step size from decreasing indefinitely with vanishing physical support of cut elements for Neumann problems [40]. However, for typical stabilization values the critical time step size still reduces to roughly one tenth of equivalent uncut elements. Moreover, for Dirichlet conditions weakly imposed via penalty or Nitsche formulations, the critical time step size can decrease without bounds.

In [41, 42], a simple yet efficient stabilization technique, referred to as eigenvalue stabilization (EVS), was proposed. Based on an eigenvalue decomposition of the elemental stiffness matrix, EVS acts locally at the element level, incurring minimal computational cost and enabling straightforward parallelization. The key concept is to identify modes associated with small or vanishing eigenvalues, group them, and construct a corresponding stabilization matrix. Since the stabilization matrices are computed during element assembly, the flexibility of the formulation is preserved and an integration into high-performance computing environments is readily possible. Originally, EVS was applied in extended finite element methods (XFEM) to mitigate ill-conditioning in enriched elements for quasi-static and dynamic crack propagation problems.

Its extension to immersed boundary methods, particularly the finite cell method (FCM), was demonstrated in [4], where EVS reduced the condition number of cut elements without degrading the achievable accuracy. For nonlinear analyses, such as hyperelasticity at finite strains, an iteratively updated force correction term was introduced [4] and later combined with remeshing to enhance robustness [6]. It is important to stress again that the overall numerical overhead remains low since the eigenvalue decomposition is restricted to cut elements and the iterative correction naturally integrates into standard nonlinear solvers.

In [5], EVS was extended to dynamics for immersed boundary finite element methods, achieving substantial condition number reductions and, for explicit dynamics, significant increases in the critical time step size of cut elements. In this context, only the mass matrix was stabilized after testing various alternative formulations. However, the nonlinear force correction scheme could not be transferred to dynamics, as altering the amplification matrix and load operator of the involved time integration schemes can cause unconditional instability. From the obtained results, we can conclude that without stabilization, critical modes can severely compromise the robustness of the FCM, especially for badly cut elements and high-order shape functions. Hence, a targeted stabilization of selected stiffness and/or mass matrix components is essential for maintaining stability and accuracy.

Considering the generalized eigenvalue problem, related approaches have been developed in the context of boundary-conforming FEM, both at the element level and at the global system level [43–45]. We extend these concepts to immersed boundary discretizations and propose a generalized eigenvalue stabilization (GEVS) strategy that aims to improve the critical time step size of the global system. In our formulation, GEVS is applied *locally* to each cut element, requiring only element-level information. Additional mild material stabilization ensures that the generalized eigenvalue problem remains well defined. However, by design, GEVS can also be combined with other basis functions and stabilization approaches. We numerically demonstrate the following aspects:

1. GEVS successfully suppresses spectral outliers in the eigenvalue spectrum, while preserving high accuracy in the remaining part.
2. In $h$-convergence studies, GEVS preserves optimal convergence rates for wave propagation problems.
3. The critical time step size is rendered independent of the element's cut ratio and geometric configuration.
4. All observations hold true for Neumann problems as well as for Dirichlet boundary conditions weakly imposed via Nitsche's method or penalty formulations.

The remaining paper is divided into four parts: In Section 2, we present the methodology, introducing the underlying problem and its spatial and temporal discretization, followed by the theoretical background of GEVS. Section 3 compares the global accuracy properties of GEVS with material and regular eigenvalue stabilization (EVS) employing a one-dimensional immersed bar problem. By means of this example, it is shown that GEVS simultaneously preserves high accuracy in the lower part of the eigenvalue spectrum and removes all outliers in the upper part. Additional studies on the asymptotic accuracy and the critical time step size for decreasing cut ratios further demonstrate that GEVS conserves optimal convergence rates and restores critical time step sizes of corresponding uncut elements. In Section 4, we validate these findings through a two-dimensional asymptotic accuracy analysis considering an immersed arc. Section 5 summarizes and concludes this contribution.

## 2. Methodology

**Model problem**

We consider a wave propagation problem underlying the second-order scalar wave equation on the domain $\Omega \subset \mathbb{R}^d$ for time $[0, T]$, where $d$ is the number of spatial dimensions and $T > 0$. The boundary $\Gamma$ of the domain $\Omega$ is divided into complementary parts $\Gamma_\text{D}$, on which Dirichlet boundary conditions $g_\text{D}$ are defined, and $\Gamma_\text{N}$, on which Neumann boundary conditions $g_\text{N}$ are imposed. Note that, as is typical, the normal vector $\boldsymbol{n}$ points outward at $\Gamma$. The initial state is defined by $\Psi_0(\boldsymbol{x})$ and $\dot{\Psi}_0(\boldsymbol{x})$. $\dot{\square}$ and $\ddot{\square}$ denote the first and second derivatives w.r.t. time. For the mass density $\rho$, the wave speed $c$, and the volumetric force $f$, the scalar wave field $\Psi : \Omega \to \mathbb{R}$ satisfies the following strong form

$$\text{(strong)} \begin{cases} \rho\, \ddot{\Psi} - \nabla \cdot \left(\rho\, c^2\, \nabla \Psi\right) = f & \text{on } \Omega \times (0, T] \\ \Psi = g_\text{D}(t) & \boldsymbol{x} \in \Gamma_\text{D}, t \in (0, T] \\ \nabla \Psi \cdot \boldsymbol{n} = g_\text{N}(t) & \boldsymbol{x} \in \Gamma_\text{N}, t \in (0, T] \\ \Psi(\boldsymbol{x}, 0) = \Psi_0(\boldsymbol{x}) & \boldsymbol{x} \in \Omega, t = 0 \\ \dot{\Psi}(\boldsymbol{x}, 0) = \dot{\Psi}_0(\boldsymbol{x}) & \boldsymbol{x} \in \Omega, t = 0 \end{cases} \quad (1)$$

Figure 1a illustrates an exemplary model domain including its Dirichlet and Neumann boundary conditions. Moreover, the boundary-conforming and non-conforming spatial discretizations are depicted in Figure 1b and Figure 1c, respectively. Note that additional information on the extended domain is also provided in Figure 1d.

For test functions $v$, the corresponding weak form reads [46, 21]:

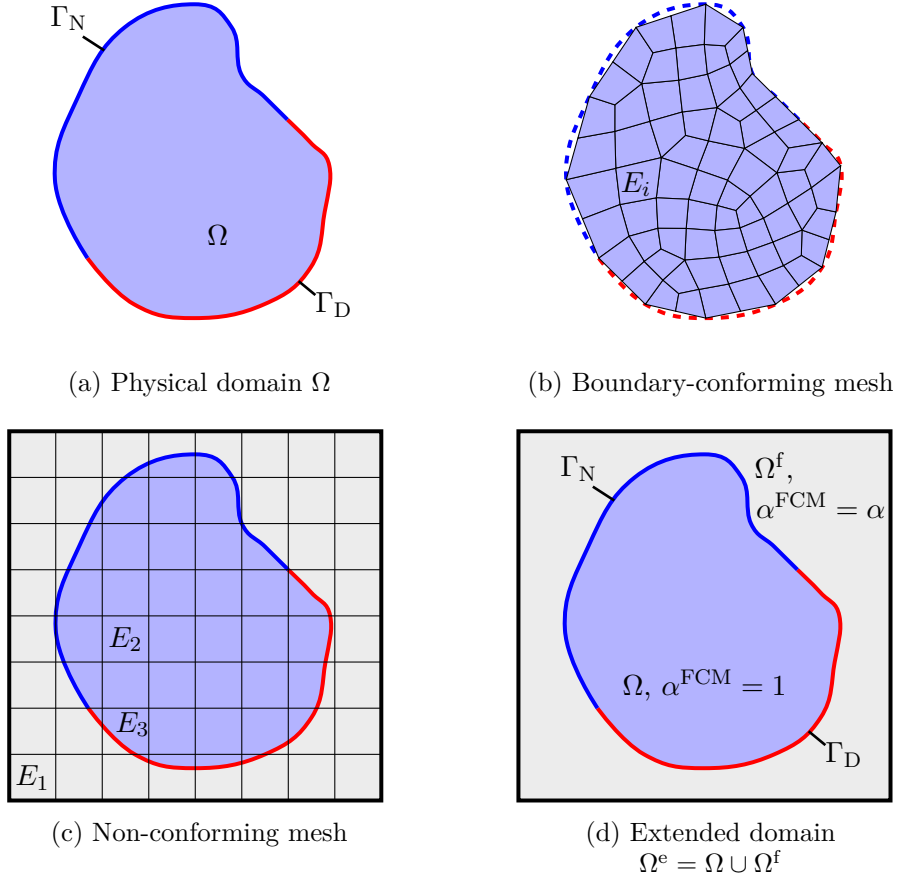

(a) Physical domain $\Omega$

(b) Boundary-conforming mesh

(c) Non-conforming mesh

(d) Extended domain $\Omega^{\mathrm{e}} = \Omega \cup \Omega^{\mathrm{f}}$

Figure 1: Schematic illustration of boundary-fitted and immersed discretizations.

Find $\Psi \in H^1_{g_{\mathrm{D}}}(\Omega)$ and $\ddot{\Psi} \in H^{-1}(\Omega)$ s.t. $\forall v \in H^1_0(\Omega)$ for a.e. $t \in (0, T]$

$$
\text{(weak)} \begin{cases} \displaystyle\int_\Omega \rho\, \ddot{\Psi}\, v\, \mathrm{d}\Omega + \int_\Omega \rho\, c^2\, \nabla \Psi \cdot \nabla v\, \mathrm{d}\Omega = \int_\Omega f\, v\, \mathrm{d}\Omega - \int_{\Gamma_{\mathrm{N}}} g_{\mathrm{N}}\, v\, \mathrm{d}\Gamma \\ \Psi(\boldsymbol{x}, 0) = \Psi_0(\boldsymbol{x}) & \text{on } \Omega \\ \dot{\Psi}(\boldsymbol{x}, 0) = \dot{\Psi}_0(\boldsymbol{x}) & \text{on } \Omega \end{cases}, \quad (2)
$$

where $H^1_{g_{\mathrm{D}}}(\Omega) = \{\Psi \in H^1(\Omega) \colon \Psi|_{\Gamma_{\mathrm{D}}} = g_{\mathrm{D}}\}$ and $H^{-1}(\Omega)$ is its dual space.

**Mesh and basis functions**

In boundary-fitted FEM, we partition the domain $\Omega$ into elements $E$, see Figure 1b. The collection of all elements forms the mesh $\mathcal{T}_h$, which covers the domain $\Omega_h$

$$\Omega_h = \bigcup_{E \in \mathcal{T}_h} E. \quad (3)$$

The computational domain $\Omega_h$ typically only approximates the original domain $\Omega$, resulting in a geometry approximation error. The global mesh size is defined as $h = \max_{E \in \mathcal{T}_h} h_E$, where the size of an element $E$ is denoted as $h_E$. On the mesh $\mathcal{T}_h$, we define the finite dimensional function space

$$V^h = \left\{ v \in C^0(\Omega_h) \colon v|_E \in N^p(E), E \in \mathcal{T}_h \right\}, \quad (4)$$

consisting of piecewise continuous multi-variate Lagrange polynomials $N^p$ of order $p$. Boundary-fitted FEM usually strongly imposes homogeneous Dirichlet boundary conditions by discarding basis functions with support on $\Gamma_D$. The remaining functions span $V_0^h \subset H_0^1(\Omega_h)$. Inhomogeneous Dirichlet boundary conditions can be imposed through lifting functions $\ell(g_D) \in H^1(\Omega_h)$, resulting in the function space $V_{g_D}^h$

$$V_{g_D}^h = \{v \in V^h : v|_{\Gamma_D} = g_D\}. \tag{5}$$

In the immersed setting, we extend the physical domain $\Omega$ by a fictitious domain $\Omega^{\mathrm{f}}$ to obtain the extended domain $\Omega^{\mathrm{e}} = \Omega \cup \Omega^{\mathrm{f}}$, see Figure 1d. The indicator function $\alpha^{\mathrm{FCM}}$ distinguishes between physical and fictitious domains

$$\alpha^{\mathrm{FCM}}(\boldsymbol{x}) = \begin{cases} 1 & \text{for } \boldsymbol{x} \in \Omega \\ \alpha & \text{else} \end{cases}, \tag{6}$$

where $\alpha \in [0,1]$ is the FCM parameter, typically chosen as a small positive number (e.g., $10^{-6}$). The extended domain $\Omega^{\mathrm{e}}$ is partitioned into elements $E$, see Figure 1c, yielding the extended mesh $\mathcal{T}_h^e$. We denote the active mesh $\tilde{\mathcal{T}}_h \subseteq \mathcal{T}_h^e$ as the set of all elements intersecting the physical domain $\Omega$, viz.

$$\tilde{\mathcal{T}}_h = \{E \in \mathcal{T}_h^e | E \cap \Omega \neq \emptyset\}. \tag{7}$$

The active domain covered by these elements is

$$\tilde{\Omega}_h = \bigcup_{E \in \tilde{\mathcal{T}}_h} E \supseteq \Omega. \tag{8}$$

Analogous to the boundary-fitted FEM, the solution space $\tilde{V}^h$ is spanned by basis functions introduced on the active mesh $\tilde{\mathcal{T}}_h$

$$\tilde{V}^h = \left\{v \in C^0(\tilde{\Omega}_h) : v|_E \in N^p(E), E \in \tilde{\mathcal{T}}_h\right\}. \tag{9}$$

All elements intersecting the boundary of the domain $\Gamma$ form the set of cut elements

$$\tilde{\mathcal{T}}_h^{\mathrm{cut}} = \left\{E \in \tilde{\mathcal{T}}_h | E \cap \Gamma \neq \emptyset\right\}, \tag{10}$$

while all elements that are fully supported on the physical domain $\Omega$ are called internal elements

$$\tilde{\mathcal{T}}_h^{\mathrm{int}} = \left\{E \in \tilde{\mathcal{T}}_h | E \subset \Omega\right\}. \tag{11}$$

In contrast to the boundary-fitted FEM, strongly imposing Dirichlet boundary conditions is not straightforward due to the mismatch between the boundary of the mesh and the physical domain. Instead, they have to be imposed in a weak sense as shown below. To this end, no constraints are included in the introduced trial and test function spaces.

The basis functions of the discrete spaces $V^h$ and $\tilde{V}^h$, defined on the meshes $\mathcal{T}_h$ and $\tilde{\mathcal{T}}_h$, respectively, are constructed via tensor products of uni-variate Lagrange polynomials. On a reference interval $[-1,1]$, we denote $\xi_{0,k}^{\mathrm{Lo},p-1}$, $k = 1, 2, \ldots, p-1$ as the roots of Lobatto polynomials of order $p-1$. Thus, the corresponding $p+1$ GLL points are defined as:

$$\xi_j = \begin{cases} -1 & \text{if } j = 1 \\ \xi_{0,j-1}^{\mathrm{Lo},p-1} & \text{if } 2 \leq j < p+1 \\ 1 & \text{if } j = p+1 \end{cases}. \tag{12}$$

Using these GLL points as interpolation nodes [29–31], the uni-variate Lagrange shape functions are defined for $\xi \in [-1,1]$ by

$$N_i^{\mathrm{Lag},p}(\xi) = \prod_{j=1, j \neq i}^{p+1} \frac{\xi - \xi_j}{\xi_i - \xi_j} \quad \text{for } i = 1, 2, \ldots, p+1. \tag{13}$$

Finally, we obtain the $(p+1)^d$ tensor product shape functions on a $d$-dimensional reference element $\boldsymbol{\xi} \in [-1, 1]^d$

$$N^p_{i_1,\ldots,i_d}(\boldsymbol{\xi}) = \prod_{k=1}^{d} N^{\text{Lag},p}_{i_k}(\xi_k) \tag{14}$$

for $i_1, \ldots, i_d = 1, \ldots, p+1$. The basis functions $N_i^p(\boldsymbol{x})$ are then obtained by mapping $N^p_{i_1,\ldots,i_d}(\boldsymbol{\xi})$ to the boundary-conforming or immersed meshes $\mathcal{T}_h$ or $\tilde{\mathcal{T}}_h$ and connecting the shape functions corresponding to boundary nodes across the element boundaries. Finally, $N_i(\boldsymbol{x})$ $i = 1, \ldots, n^{\text{dof}}$ span the approximation space $V^h$ or $\tilde{V}^h$, where $n^{\text{dof}}$ denotes the total number of active degrees of freedom of the respective discretizations.

Assuming that the functions evolve smoothly in time from $t = 0$ to $t = T$, the semi-discrete boundary-conforming and immersed function spaces $V^h_{g_D,T}$, $V^h_T$, and $\tilde{V}^h_T$ are [21]:

$$V^h_{g_D,T} = V^h_{g_D} \otimes \mathcal{C}^\infty \tag{15}$$
$$V^h_T = V^h \otimes \mathcal{C}^\infty \tag{16}$$
$$\tilde{V}^h_T = \tilde{V}^h \otimes \mathcal{C}^\infty. \tag{17}$$

**Spatial and temporal discretization**

Considering boundary-conforming FEM, any test function $v^h \in V^h_0$ and trial function $\Psi^h \in V^h_{g_D,T}$ can be represented as a linear combination of the basis functions

$$v^h(\boldsymbol{x}) = \sum_{i=1}^{n^{\text{dof},0}} N_i(\boldsymbol{x})\, \hat{v}_i = \mathbf{N}(\boldsymbol{x})\hat{\boldsymbol{v}} \tag{18}$$

$$\Psi^h(\boldsymbol{x}, t) = \sum_{i=1}^{n^{\text{dof},0}} N_i(\boldsymbol{x})\, \hat{\Psi}_i(t) + \ell(g_D) = \mathbf{N}(\boldsymbol{x})\hat{\boldsymbol{\Psi}}(t) + \ell(g_D) \tag{19}$$

with arbitrary test coefficients $\hat{v}_i$ and time-dependent trial coefficients $\hat{\Psi}_i(t) \in \mathcal{C}^\infty(0,T)$, where $n^{\text{dof},0}$ is the number of basis functions active in $V^h_0$. We arrive at the semi-discrete finite element formulation

Find $\Psi^h \in V^h_{g_D,T}$ s.t. $\forall v^h \in V^h_0$

$$(\text{boundary-fitted}) \begin{cases} \int_{\Omega_h} \rho\, \ddot{\Psi}^h\, v^h\, \mathrm{d}\Omega + \int_{\Omega_h} \rho\, c^2\, \nabla \Psi^h \cdot \nabla v^h\, \mathrm{d}\Omega \\ \qquad = \int_{\Omega_h} f\, v^h\, \mathrm{d}\Omega - \int_{\Gamma^{\text{N}}} g_{\text{N}}\, v^h\, \mathrm{d}\Gamma \end{cases}. \tag{20}$$

Using the representation of trial and test functions (19), we can compute the element mass matrix $\boldsymbol{m}^E$, stiffness matrix $\boldsymbol{k}^E$, and force vector $\boldsymbol{f}^E$

$$[\boldsymbol{m}^E]_{ij} = \int_E \rho\, N_i(\boldsymbol{x})\, N_j(\boldsymbol{x})\, \mathrm{d}\Omega \tag{21}$$

$$[\boldsymbol{k}^E]_{ij} = \int_E \rho\, c^2\, \nabla N_i(\boldsymbol{x}) \cdot \nabla N_j(\boldsymbol{x})\, \mathrm{d}\Omega \tag{22}$$

$$[\boldsymbol{f}^E]_i = \int_E f(\boldsymbol{x})\, N_i(\boldsymbol{x})\, \mathrm{d}\Omega + \int_{\Gamma^E_N} g_{\text{N}}(\boldsymbol{x})\, N_i(\boldsymbol{x})\, \mathrm{d}\Gamma \tag{23}$$

for $E \in \mathcal{T}_h$, where $i, j = 1, \ldots, n^{\text{dof},E}$ with $n^{\text{dof},E}$ being the number of basis functions active in element $E$, and $\Gamma^E_N = E \cap \Gamma_{\text{N}}$ is the element's part of the Neumann boundary.

Considering immersed boundary FEM, the test functions $v^h \in \tilde{V}^h$ and trial function $\Psi^h \in \tilde{V}_T^h$ are analogously defined as a linear combination of the corresponding basis functions

$$v^h(\boldsymbol{x}) = \sum_{i=1}^{n^{\text{dof}}} N_i(\boldsymbol{x})\,\hat{v}_i = \mathbf{N}(\boldsymbol{x})\hat{\boldsymbol{v}} \tag{24}$$

$$\Psi^h(\boldsymbol{x},t) = \sum_{i=1}^{n^{\text{dof}}} N_i(\boldsymbol{x})\,\hat{\Psi}_i(t) = \mathbf{N}(\boldsymbol{x})\hat{\boldsymbol{\Psi}}(t) \tag{25}$$

with the coefficients $\hat{v}_i$ and $\hat{\Psi}_i(t) \in \mathcal{C}^\infty(0,T)$. By applying the symmetric version of Nitsche's method, the immersed semi-discrete finite element formulation reads as follows [47]:

Find $\Psi^h \in \tilde{V}_T^h$ s.t. $\forall v^h \in \tilde{V}^h$

$$\text{(immersed)} \begin{cases} \int_{\tilde{\Omega}_h} \alpha^{\text{FCM}}\,\rho\,\ddot{\Psi}^h\,v^h\,\mathrm{d}\Omega + \int_{\tilde{\Omega}_h} \alpha^{\text{FCM}}\,\rho\,c^2\,\nabla\Psi^h \cdot \nabla v^h\,\mathrm{d}\Omega \\ \underbrace{- \int_{\Gamma_\mathrm{D}} \partial_n \Psi^h\,v^h\,\mathrm{d}\Gamma - \int_{\Gamma_\mathrm{D}} \partial_n v^h\,\Psi^h\,\mathrm{d}\Gamma + \int_{\Gamma_\mathrm{D}} \lambda_E\,\Psi^h\,v^h\,\mathrm{d}\Gamma}_{\text{Nitsche terms}} \\ = \int_{\tilde{\Omega}_h} f\,v^h\,\mathrm{d}\Omega + \int_{\Gamma_\mathrm{N}} g_\mathrm{N}\,v^h\,\mathrm{d}\Gamma \underbrace{- \int_{\Gamma_\mathrm{D}} \partial_n v^h\,g_\mathrm{D}\,\mathrm{d}\Gamma + \int_{\Gamma_\mathrm{D}} \lambda_E\,g_\mathrm{D}\,v^h\mathrm{d}\Gamma}_{\text{Nitsche terms}} \end{cases}. \tag{26}$$

where $\partial_n = \boldsymbol{n} \cdot \nabla$ is the derivative in normal direction on $\Gamma$ and $\lambda_E$ is the element-wise Nitsche parameter. Note that the volume integrals are taken over the active domain $\tilde{\Omega}_h$. To ensure coercivity of the bilinear form on the left-hand side of the above equation, the penalty parameter $\lambda_E$ has to be sufficiently large [47]. As suggested in [48–50, 47], we solve the local/element-wise eigenvalue problem:

Find $(\Psi^h_\mu, \mu_E) \in \tilde{V}^h|_E^0 \times \mathbb{R}$, such that

$$\int_{\Gamma_\mathrm{D}^E} \partial_n \Psi^h_\mu\,\partial_n v^h\,\mathrm{d}\Gamma = \mu_E \int_{\Omega^E} \nabla \Psi^h_\mu \cdot \nabla v^h \mathrm{d}\Gamma \quad \text{for all } v^h \in \tilde{V}^h|_E^0 \tag{27}$$

for all elements $E$ which intersect with the Dirichlet boundary $\Gamma_\mathrm{D}$ and with $\Gamma_\mathrm{D}^E = E \cap \Gamma_\mathrm{D}$ denoting the Dirichlet boundary contribution of element $E$ and its physical part $\Omega^E = E \cap \Omega$. In [47, 48], it is suggested to select the penalty parameter to be twice the largest eigenvalue

$$\lambda_E = 2 \max \mu_E. \tag{28}$$

Now, using the representation of trial and test functions for the immersed configuration (25), the element mass matrix $\boldsymbol{m}^E$, stiffness matrix $\boldsymbol{k}^E$, and force vector $\boldsymbol{f}^E$ are:

$$[\boldsymbol{m}^E]_{ij} = \int_E \rho\,N_i(\boldsymbol{x})\,N_j(\boldsymbol{x})\,\mathrm{d}\Omega \tag{29}$$

$$[\boldsymbol{k}^E]_{ij} = \int_E \rho\,c^2\,\nabla N_i(\boldsymbol{x}) \cdot \nabla N_j(\boldsymbol{x})\,\mathrm{d}\Omega \tag{30}$$

$$- \int_{\Gamma_\mathrm{D}^E} \partial_n N_i(\boldsymbol{x})\,N_j(\boldsymbol{x})\mathrm{d}\Gamma - \int_{\Gamma_\mathrm{D}^E} N_i(\boldsymbol{x})\,\partial_n N_j(\boldsymbol{x})\mathrm{d}\Gamma + \int_{\Gamma_\mathrm{D}^E} \lambda_E\,N_i(\boldsymbol{x})\,N_j(\boldsymbol{x})\mathrm{d}\Gamma$$

$$[\boldsymbol{f}^E]_i = \int_E f(\boldsymbol{x})\,N_i(\boldsymbol{x})\,\mathrm{d}\Omega + \int_{\Gamma_\mathrm{N}^E} g_\mathrm{N}(\boldsymbol{x})\,N_i(\boldsymbol{x})\,\mathrm{d}\Gamma \tag{31}$$

$$- \int_{\Gamma_\mathrm{D}^E} g_\mathrm{D}(\boldsymbol{x})\,\partial_n N_i(\boldsymbol{x})\mathrm{d}\Gamma + \int_{\Gamma_\mathrm{D}^E} \lambda_E\,g_\mathrm{D}(\boldsymbol{x})\,N_i(\boldsymbol{x})\mathrm{d}\Gamma$$

for $E \in \tilde{\mathcal{T}}_h$, where $i, j = 1, \ldots, n^{\text{dof},E}$ with $n^{\text{dof},E}$ being the number of basis functions active in element $E$, and $\Gamma_{\text{N}}^E = E \cap \Gamma_{\text{N}}$ and $\Gamma_{\text{D}}^E = E \cap \Gamma_{\text{D}}$ is the element's part of the Neumann and Dirichlet boundary.

For either the boundary conforming case (21) to (23) or immersed case (29) to (31), we obtain the global mass and stiffness matrices $\mathbf{M}$ and $\mathbf{K}$ and the force vector $\mathbf{F}$ after assembly of the elements

$$\mathbf{M} = \underset{E \in \tilde{\mathcal{T}}_h}{\mathcal{A}} \boldsymbol{m}^E, \; \mathbf{K} = \underset{E \in \tilde{\mathcal{T}}_h}{\mathcal{A}} \boldsymbol{k}^E, \text{ and } \mathbf{F} = \underset{E \in \tilde{\mathcal{T}}_h}{\mathcal{A}} \boldsymbol{f}^E. \tag{32}$$

where $\mathcal{A}$ denotes the standard assembly operation over all active elements.

The global semi-discrete system is

$$\mathbf{M}\ddot{\hat{\boldsymbol{\Psi}}} + \mathbf{K}\hat{\boldsymbol{\Psi}} = \mathbf{F}. \tag{33}$$

To integrate the semi-discrete system in time, we employ the explicit central difference method (CDM) with second-order accuracy

$$\hat{\boldsymbol{\Psi}}_{k+1} = 2\,\hat{\boldsymbol{\Psi}}_k - \hat{\boldsymbol{\Psi}}_{k-1} + \Delta t^2 \mathbf{M}^{-1} \left( \mathbf{F}_k - \mathbf{K}\,\hat{\boldsymbol{\Psi}}_k \right). \tag{34}$$

Its stability is constrained by the critical time step size

$$\Delta t^{\text{crit}} = \frac{2}{\sqrt{\lambda_{\max}(\mathbf{K}, \mathbf{M})}}, \tag{35}$$

where $\lambda_{\max}(\mathbf{K}, \mathbf{M})$ denotes the largest eigenvalue of the generalized eigenvalue problem

$$\mathbf{K}\hat{\boldsymbol{\phi}} = \lambda \mathbf{M}\hat{\boldsymbol{\phi}}. \tag{36}$$

**Material stabilization**

To distinguish between the physical and fictitious domains, the indicator function $\alpha^{\text{FCM}}$ – see (6) – is introduced. It also serves a second purpose, that is to stabilize cut elements. Selecting a small but positive value of $\alpha$ ensures that the system remains well-posed. Within the fictitious domain, a nonzero $\alpha$ corresponds to adding very soft material, which limits the attainable accuracy by introducing a consistency error into the formulation. For Laplace and Helmholtz problems, Dauge et al. [51] showed that the corresponding error in the energy norm scales as $\sqrt{\alpha}$. In practice, $\alpha$ is typically still chosen between $10^{-20}$ and $10^{-2}$ to improve the conditioning and stability of the discretized system. In static problems, $\alpha$ primarily affects the conditioning of the stiffness matrix, thereby accelerating the convergence of iterative solvers. On the other hand, in dynamic problems, it influences the largest eigenvalue of the generalized eigenvalue problem – see (36) – and thus governs the critical time step size in explicit time integration schemes. Note that a detailed discussion of material stabilization and its effect on the critical time step can be found in [40].

**Regular eigenvalue stabilization**

In [4], Garhuom et al. introduce an eigenvalue stabilization technique for nonlinear statics with immersed boundary finite element discretizations. Cut elements, especially when they exhibit small volume fractions, lead to severe ill-conditioning of the global stiffness matrix. Adding additional stiffness based on the dyadic product of the eigenvectors corresponding to small eigenvalues alleviates this issue. Note that this technique has been pioneered by Löhnert in the context of the extended FEM (XFEM) [41, 42], where similar problems arise. Considering dynamics and wave propagation analysis, Eisenträger et al. [5] present a related strategy to address the adverse effects of cut elements on the critical time step size. They stabilize the smallest eigenvalues of the mass matrices at the element level to restore stable explicit time integration. In the following, we denote the regular eigenvalue stabilization as EVS.

For all cut elements $E \in \tilde{\mathcal{T}}_h^{\text{cut}}$, the stabilized mass matrix $\tilde{\boldsymbol{m}}^E$ is a sum of the original mass matrix $\boldsymbol{m}^E$ and a stabilization term $\boldsymbol{m}^{\text{s}}$ multiplied by the stabilization parameter $\epsilon$:

$$\tilde{\boldsymbol{m}}^E = \boldsymbol{m}^E + \epsilon\,\boldsymbol{m}^{\text{s}}. \tag{37}$$

The set of eigenpairs of the unstabilized mass matrix $\{(\lambda_i, \phi_i)\}$ contains the eigenvalues and corresponding eigenvectors $\lambda_i$ and $\phi_i$ for $i = 1, \ldots, n^{\mathrm{dof},E}$. In the following, we denote the set of eigenpairs with small eigenvalues as $\Phi^{\mathrm{s}} = \{(\lambda_i, \phi_i) \,|\, \lambda_i < f_\lambda \lambda_{\max}\}$, where $f_\lambda$ is a chosen treshold and $\lambda_{\max}$ is the largest eigenvalue of the element. The unscaled stabilization term is

$$\boldsymbol{m}^{\mathrm{s},0} = \sum_{(\lambda_i, \phi_i) \in \Phi^{\mathrm{s}}} \phi_i \phi_i^T. \tag{38}$$

In Appendix A, we show that adding this term to the original mass matrix just manipulates the eigenvalues in $\Phi^{\mathrm{s}}$, while the rest of the spectrum remains unchanged. The final stabilization matrix (independent of the chosen material and element geometry) is

$$\boldsymbol{m}^{\mathrm{s}} = \frac{\max(\boldsymbol{m}^{\mathrm{f}})}{\max(\boldsymbol{m}^{\mathrm{s},0})} \boldsymbol{m}^{\mathrm{s},0}, \tag{39}$$

where $\boldsymbol{m}^{\mathrm{f}}$ is the mass matrix of a full or internal element and $\max(\mathbf{M})$ denotes the maximum element of a matrix $\mathbf{A}$. In the following, we set the threshold parameter $f_\lambda$ to a value of $10^{-2}$. The stabilization parameter $\epsilon$ is varied to achieve an appropriate amount of stabilization. The stabilized global mass matrix is

$$\tilde{\mathbf{M}} = \underset{E \in \tilde{\mathcal{T}}_h^{\mathrm{int}}}{\mathcal{A}} \boldsymbol{m}^E + \underset{E \in \tilde{\mathcal{T}}_h^{\mathrm{cut}}}{\mathcal{A}} \tilde{\boldsymbol{m}}^E. \tag{40}$$

### 2.1. Generalized eigenvalue stabilization

In the context of stabilization methods, it is important to note that regular eigenvalue stabilization techniques modify the entire spectrum of an element's generalized eigenvalue problem, thereby modifying also the low-frequency modes that are essential for accuracy. A more selective strategy can be achieved through deflation strategies applied directly to the generalized eigenvalue problem [52]. In boundary-conforming FEM, Tkachuk et al. [43] introduced such an approach under the name spectral mass scaling to mitigate unwanted high frequencies, and investigated its efficiency both at the local element and global system levels. This approach was further refined by González and Park [53], who introduced mesh partitioning and an efficient solution strategy based on the Woodbury matrix identity. They refer to their approach as mass tailoring. Moreover, Voet et al. [45, 54] demonstrated that a generalization of this deflation strategy is also suitable for outlier removal in IGA at the global system level. In this work, we extend this concept for immersed-boundary discretizations and apply rank-one modifications on the generalized eigenvalue problem for cut elements locally at element level. This allows us to selectively stabilize the largest eigenvalues in cut elements without altering the rest of the spectrum, see Appendix B. In the following, we refer to this approach as generalized eigenvalue stabilization (GEVS). We consider three variants of GEVS:

(i) Deflation of the mass matrix.
(ii) Deflation of the stiffness matrix.
(iii) Deflation of mass and stiffness matrices.

(i) *GEVS by mass deflation*: Analogous to the standard EVS approach [4, 5], the stabilized mass matrix $\tilde{\boldsymbol{m}}^E$ is obtained by augmenting the original mass matrix $\boldsymbol{m}^E$ with a stabilization term $\boldsymbol{m}^{\mathrm{s}}$ comprising rank-one modifications

$$\tilde{\boldsymbol{m}}^E = \boldsymbol{m}^E + \boldsymbol{m}^{\mathrm{s}}. \tag{41}$$

Let $\{(\lambda_i, \phi_i)\}$ denote the set of eigenpairs of the element's generalized eigenvalue problem, where $\lambda_i$ and $\phi_i$ are the eigenvalues and corresponding eigenvectors, for $i = 1, \ldots, n^{\mathrm{dof},E}$. The reference eigenvalue $\lambda^*$ denotes the largest eigenvalue of the corresponding uncut element. We define the stabilization term as

$$\boldsymbol{m}^{\mathrm{s}} = \sum_{(\lambda_i, \phi_i) \in \Phi^{\mathrm{l}}} c_i \, \boldsymbol{m}^E \phi_i \phi_i^T \boldsymbol{m}^E, \tag{42}$$

where the set $\Phi^l = \{(\lambda_i, \boldsymbol{\phi}_i) \,|\, \lambda_i > \lambda^*\}$ contains all eigenpairs with eigenvalues larger then $\lambda^*$. As shown in Appendix B, we choose the coefficients

$$c_i = \frac{\lambda_i}{\lambda^*} - 1 \tag{43}$$

to shift all eigenvalues in $\Phi^l$ to $\lambda^*$, without altering the remainder of the spectrum. This reduces the largest eigenvalue of the cut element, making it equal to that of the corresponding uncut element.

(ii) *GEVS by stiffness deflation*: In contrast to the mass deflation approach, we manipulate the spectrum of the generalized eigenvalue problem by applying rank-one modifications to the element stiffness matrix. The stabilized stiffness matrix $\tilde{\boldsymbol{k}}^E$ is defined as

$$\tilde{\boldsymbol{k}}^E = \boldsymbol{k}^E + \boldsymbol{k}^s, \tag{44}$$

where the stabilization term is given by

$$\boldsymbol{k}^s = \sum_{(\lambda_i, \boldsymbol{\phi}_i) \in \Phi^l} c_i \, \boldsymbol{m}^E \boldsymbol{\phi}_i \boldsymbol{\phi}_i^T \boldsymbol{m}^E. \tag{45}$$

As before, the set $\Phi^l = \{(\lambda_i, \boldsymbol{\phi}_i) \,|\, \lambda_i > \lambda^*\}$ contains all eigenpairs with eigenvalues larger than the reference eigenvalue $\lambda^*$ of an equivalent full element. To shift each eigenvalue in $\Phi^l$ to $\lambda^*$, we select the coefficients as

$$c_i = \lambda^* - \lambda_i. \tag{46}$$

In Appendix B, we also show that this reduces the largest eigenvalue to the reference of an uncut element $\lambda^*$.

(iii) *GEVS by mass and stiffness deflation*: In this variant, we simultaneously modify the mass and stiffness matrices to stabilize the generalized eigenvalue problem. The stabilized element matrices are defined as

$$\tilde{\boldsymbol{m}}^E = \boldsymbol{m}^E + \boldsymbol{m}^s \quad \text{and} \quad \tilde{\boldsymbol{k}}^E = \boldsymbol{k}^E + \boldsymbol{k}^s. \tag{47}$$

As before, the set $\Phi^l = \{(\lambda_i, \boldsymbol{\phi}_i) \,|\, \lambda_i > \lambda^*\}$ denotes the eigenpairs with eigenvalues exceeding $\lambda^*$. The stabilization terms are expressed as

$$\boldsymbol{m}^s = \sum_{(\lambda_i, \boldsymbol{\phi}_i) \in \Phi^l} c_i^m \, \boldsymbol{m}^E \boldsymbol{\phi}_i \boldsymbol{\phi}_i^T \boldsymbol{m}^E \quad \text{and} \quad \boldsymbol{k}^s = \sum_{(\lambda_i, \boldsymbol{\phi}_i) \in \Phi^l} c_i^k \, \boldsymbol{m}^E \boldsymbol{\phi}_i \boldsymbol{\phi}_i^T \boldsymbol{m}^E. \tag{48}$$

To shift the eigenvalue $\lambda_i$ to the target value $\lambda^*$, we select the coefficients

$$c_i^m = \frac{\lambda_i + c_i^k - \lambda^*}{\lambda^*} \quad \text{and} \quad c_i^k = \frac{\lambda^* - \lambda_i}{2}. \tag{49}$$

In that way the affected eigenvalues $\lambda_i$ are shifted to $\lambda^*$ equally by deflating the mass matrix and by deflating the stiffness matrix., see Appendix B.

For the stabilization approaches i) and iii), the global mass matrix is assembled to

$$\tilde{\mathbf{M}} = \underset{E \in \tilde{\mathcal{T}}_h^{\text{int}}}{\mathcal{A}} \boldsymbol{m}^E + \underset{E \in \tilde{\mathcal{T}}_h^{\text{cut}}}{\mathcal{A}} \tilde{\boldsymbol{m}}^E. \tag{50}$$

For approaches ii) and iii), the global stiffness matrix is assembled to

$$\tilde{\mathbf{K}} = \underset{E \in \tilde{\mathcal{T}}_h^{\text{int}}}{\mathcal{A}} \boldsymbol{k}^E + \underset{E \in \tilde{\mathcal{T}}_h^{\text{cut}}}{\mathcal{A}} \tilde{\boldsymbol{k}}^E. \tag{51}$$

Note that Dirichlet boundary conditions can also be imposed weakly in uncut elements, which may severely reduce their critical time step size. In these cases, GEVS can be applied to the corresponding uncut elements to address the critical time step size. However, modifying the mass matrix will result in a loss of diagonality.

## 2.2. Note on mass lumping

From the definition of the stabilization matrices obtained via the EVS and GEVS techniques, it follows that these matrices are, by construction, fully populated. In the context of explicit dynamics, employing Lagrangian interpolation polynomials defined on GLL points as basis functions only the mass matrices of uncut elements are lumped using nodal quadrature [35, 31]. This naturally raises the question of whether it is meaningful to also diagonalize the mass matrices of the stabilized cut elements. To date, the only known mass-lumping technique that preserves the positive-definiteness of the mass matrix is the HRZ scheme [55], which, however, has been shown to reduce the attainable convergence rates [56, 28]. Moreover, in immersed settings, spurious oscillations have been reported [57, 25, 58], effectively rendering this approach unsuitable for practical use. We therefore conclude that mass lumping for uncut elements remains an open problem requiring further dedicated research. One potential approach is the use of approximate dual basis functions, as recently proposed in isogeometric analysis [59, 60]. However, this technique is still in its early stages and entails the drawback of producing unsymmetric and more densely populated stiffness matrices. Furthermore, its extension to immersed settings has yet to be demonstrated.

# 3. Waves in a rod

### Problem statement

First, we consider a one-dimensional immersed setup. The physical domain is defined as $\Omega = [0, l_\mathrm{p}]$ and embedded in an extended (computational) domain $\Omega^\mathrm{e} = [0, l]$, where $l_\mathrm{p} = 0.9863$ and $l = 1.0$. A homogeneous Neumann boundary condition is imposed at $x = 0$, while at $x = l_\mathrm{p}$, we consider two cases: either a homogeneous Neumann boundary condition or a Dirichlet boundary condition weakly enforced using Nitsche's method, where an optimal penalty factor is computed according to (27) and (28). The material parameters are uniform within the physical domain: $\rho = c = 1$. In the fictitious domain, we set $\rho = \alpha$ and $c = 1$ with $\alpha > 0$. The cut element is integrated exactly, by bisecting it and distributing $p+1$ Gauss-Legendre points in the physical and fictitious parts. Figure 2 illustrates the domain configuration. Note that elements not intersecting the physical domain are depicted herein, but not considered in computations.

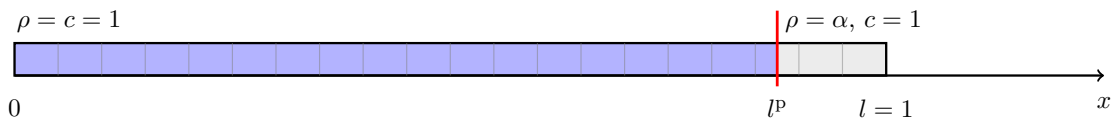

Figure 2: Investigated one-dimensional configuration.

### Global accuracy

First, we assess the global accuracy of the immersed configuration considering the ratio of all discrete eigenfrequencies $\omega_i = \sqrt{\lambda_i}$ w.r.t. their analytic reference value $\omega_i^\mathrm{ref}$. We apply material stabilization (MS), regular eigenvalue stabilization (EVS), and the proposed generalized eigenvalue stabilization (GEVS). The extended domain $\Omega^\mathrm{e}$ is discretized using 50 quadratic elements, resulting in a total of 101 degrees of freedom. Due to the position of the immersed interface at $l_\mathrm{p} = 0.9863$, only the last element of the mesh is intersected by the boundary. Figures 3, 4, and 5 show the global accuracy of MS, EVS, and GEVS, respectively, considering both Neumann and Dirichlet boundary conditions at the immersed boundary. Even in the case of EVS and GEVS, we apply a mild material stabilization ($\alpha = 10^{-10}$) to ensure that the regular and generalized eigenvalue problems of cut elements are well defined. For reference, each plot also includes the global accuracy of a corresponding boundary-conforming spectral element discretization with uniform elements. Before discussing the results of the individual stabilization approaches, we highlight several typical features related to global accuracy analyses:

- Discretizations using $C^0$-continuous basis functions exhibit so-called optical branches in the global spectrum [61, 62]. For quadratic elements, a single optical branch appears in the middle of the spectrum for both boundary-conforming and immersed discretizations.

- In boundary-conforming SEM, nodal mass lumping causes the upper part of the eigenvalue spectrum to be underestimated [63, 64]. This results in a larger critical time step size compared to standard FEM.

- In immersed methods, cut (or trimmed) elements introduce outliers at the upper end of the spectrum [58]. These outliers cause prohibitively small critical time step sizes when using explicit time integration schemes.

The different stabilization approaches are primarily assessed with respect to two criteria: the preservation of accuracy in the lower part of the eigenvalue spectrum and the suppression of outliers at its upper end.

First, we analyze the global accuracy obtained with MS (Figure 3), considering three distinct values of the stabilization parameter $\alpha = 10^{-1}$, $10^{-5}$, and $10^{-10}$. For the strongest stabilization ($\alpha = 10^{-1}$), we observe an offset already in the low eigenvalue part of the spectrum ($i = 1, \ldots, 30$) for the Neumann problem, and a significant drop in accuracy around $i = 34$ for the Dirichlet case. At the same time, the largest outlier decreases below a value of 1.2 times the corresponding analytic one. As $\alpha$ decreases, the accuracy in the lower part of the spectrum, i.e., below the optical branch, improves. For $\alpha = 10^{-5}$ and $\alpha = 10^{-10}$, the eigenfrequencies in this range closely match those of the boundary-conforming reference solution. However, in both cases, the highest eigenfrequency of the discretized system is almost three times higher compared to the analytically correct value for both Neumann and Dirichlet configurations.

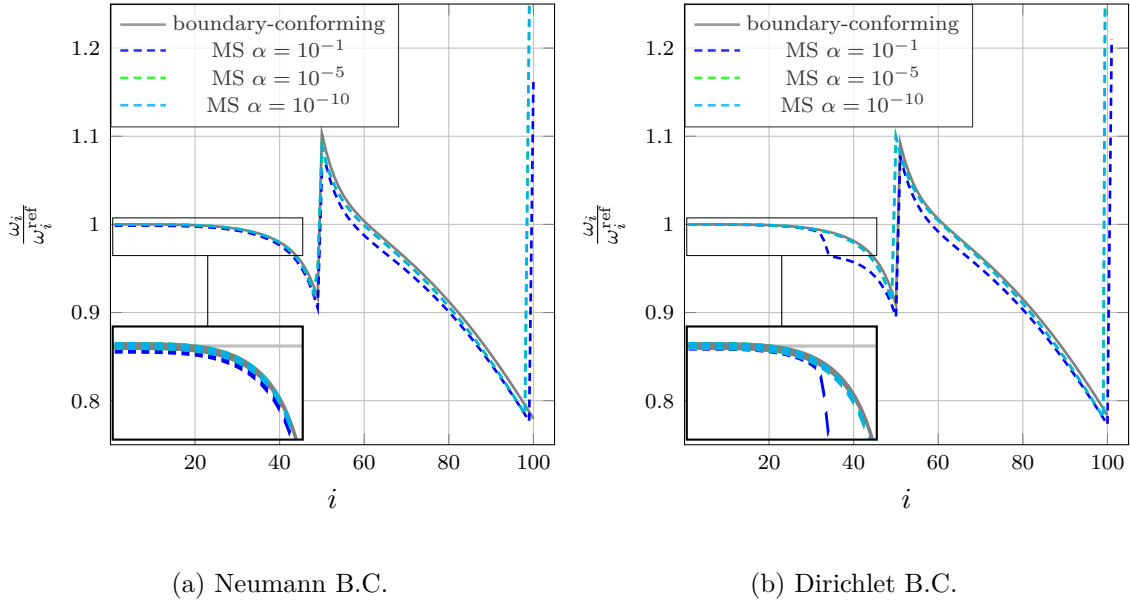

(a) Neumann B.C.   (b) Dirichlet B.C.

Figure 3: Global accuracy of MS with $n^{\text{el}} = 50$ and $p = 2$.

Next, we consider EVS (Figure 4) using $\epsilon = 10^{-2}$ and $10^{-4}$. The resulting trends and observations are qualitatively similar to those observed for MS. With strong stabilization ($\epsilon = 10^{-2}$), the spectral outlier is reduced to approximately 1.2 times the analytical value in the Neumann case, and 2.3 times in the Dirichlet case. However, an abrupt drop in the accuracy before the optical branch is induced. By contrast, mild stabilization ($\epsilon = 10^{-4}$) improves accuracy in the lower part of the eigenvalue spectrum, but fails to sufficiently suppress the outliers, which are 2.0 and 2.5 times the analytic values for Neumann and Dirichlet configurations, respectively.

For GEVS (Figure 5), we consider three variants: mass deflation, stiffness deflation, and combined mass and stiffness deflation. All three approaches succeed in effectively suppressing the spectral outlier. Equivalent to the boundary-conforming solution, the highest eigenfrequency is underestimated to approximately 0.77 times the analytic value for both Neumann and Dirichlet boundary conditions. However, deflating the stiffness matrix leads to a noticeable loss of accuracy in the lower part of the spectrum. By contrast, mass deflation maintains the accuracy across the entire spectrum and closely matches the boundary-conforming reference. The combined mass and stiffness deflation also achieves high overall accuracy but performs slightly worse than pure mass deflation. The global accuracy analysis demonstrates that GEVS with mass deflation is the only approach that successfully suppresses outliers while maintaining high accuracy in the entire spectrum. Based on these results, we focus on this variant, which we just refer to as GEVS in the remainder of this paper. In the following, we investigate its asymptotic accuracy and robustness with respect to small cuts, before applying it to a two-dimensional example in Section 4.

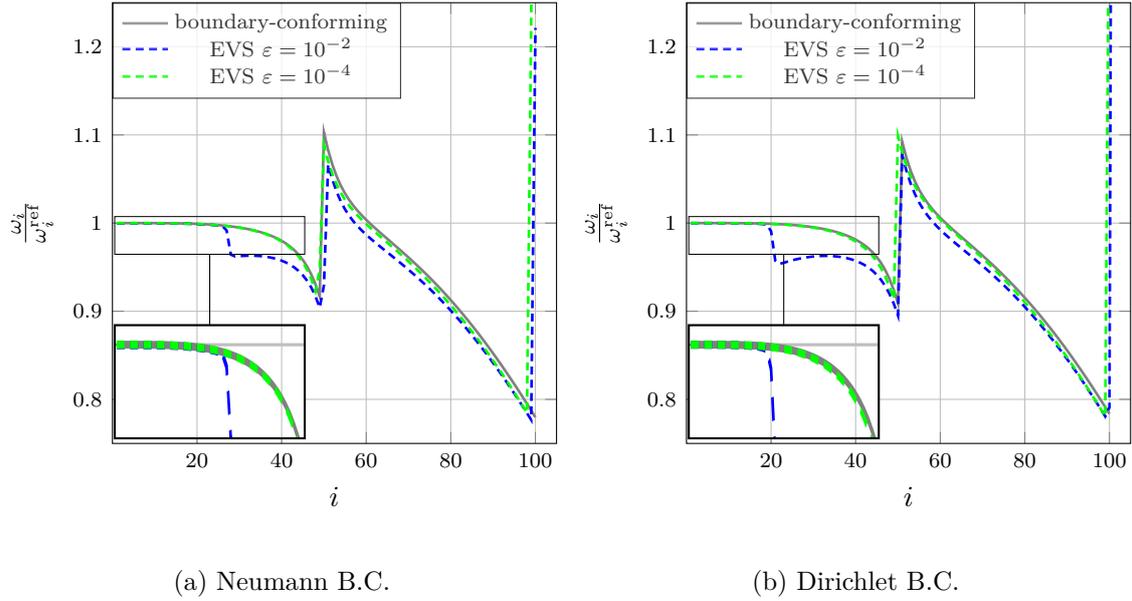

(a) Neumann B.C.  (b) Dirichlet B.C.

Figure 4: Global accuracy of EVS with $n^{el} = 50$ and $p = 2$.

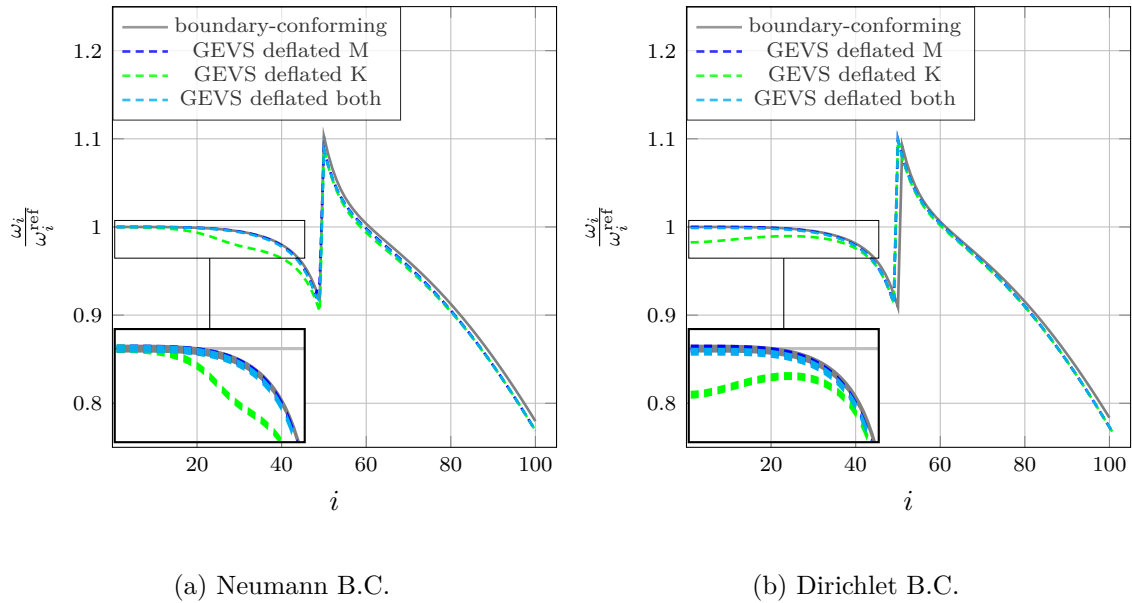

(a) Neumann B.C.  (b) Dirichlet B.C.

Figure 5: Global accuracy of GEVS with $n^{el} = 50$ and $p = 2$.

**Asymptotic accuracy**

We investigate the asymptotic accuracy of GEVS through an $h$-convergence study for polynomial orders $p = 1$ to $p = 4$. The number of elements is sampled logarithmically as $n^{\text{el}} = [10, 20, 40, 80]$. Elements that are entirely located in the fictitious domain are excluded from the computational mesh. The immersed interface remains at $l_{\text{p}} = 0.9863$, which leads to varying cut configurations of the intersected element as the mesh is refined. The initial condition is a Gaussian pulse centered on the left side of the domain:

$$\Psi_0(x) = 2\, e^{\frac{-x^2}{2\sigma^2}} \tag{52}$$

with $\sigma = 0.05$. The wave propagates to the right, gets reflected at the boundary, and returns to the origin. At the final time $T = \frac{2\, l_{\text{p}}}{c}$, the analytic solution is $\Psi(x,T) = \Psi_0(x)$ for the Neumann problem, while it is $\Psi(x,T) = -\Psi_0(x)$ for the Dirichlet problem. Time integration is performed using CDM with 100 000 time steps. Figures 6 and 7 present the $L_2$-error at the final time step as well as the corresponding critical time step sizes of the configurations for both Neumann and Dirichlet boundary conditions. For reference, the results of boundary-conforming discretizations are included as dashed lines, indicating the optimal convergence rates.

The asymptotic study indicates that deviations in both accuracy and critical time-step size between the immersed setting with GEVS and the boundary-conforming discretization are minor, where the observed differences can be attributed to the differences in element sizes between the two mesh types. However, the most important outcome is that GEVS maintains the optimal convergence order in the $L_2$-norm, with the error scaling as $\propto h^{p+1}$. At the same time, it successfully suppresses the spectral outliers and recovers critical time step sizes of equivalent boundary-conforming discretizations.

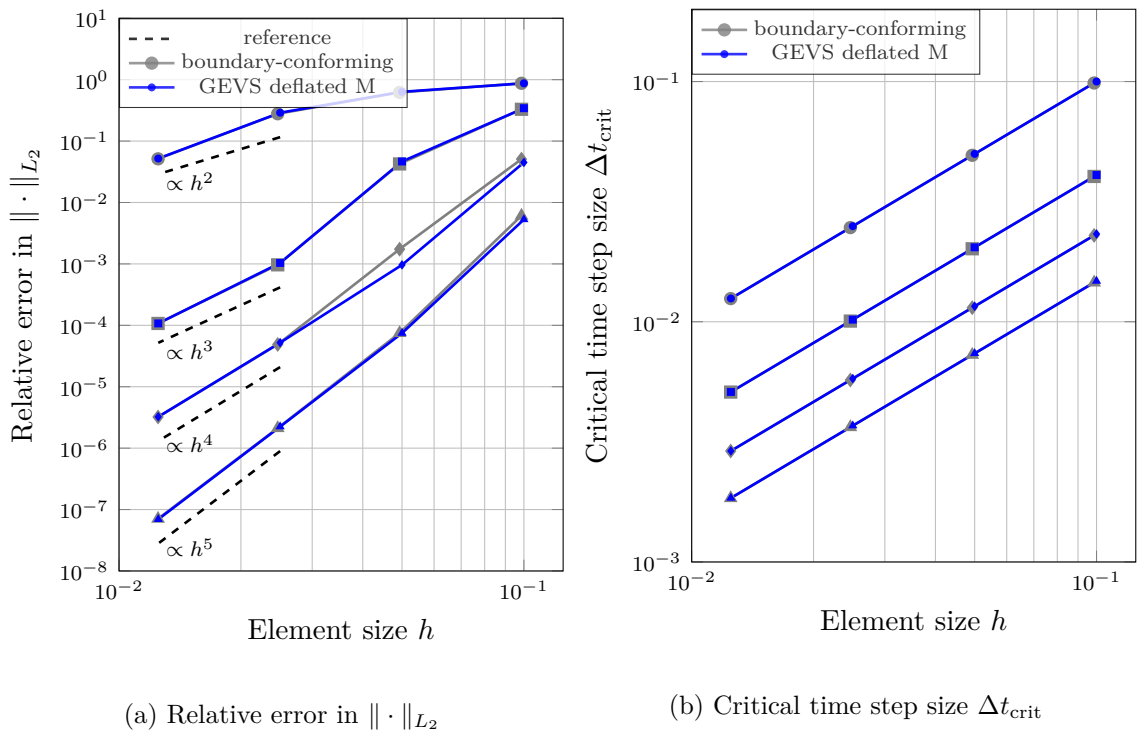

(a) Relative error in $\|\cdot\|_{L_2}$

(b) Critical time step size $\Delta t_{\text{crit}}$

Figure 6: Asymptotic accuracy and critical time step size of GEVS with Neumann boundary condition. Polynomial degrees $p = 1, \ldots, 4$ are represented by different marker shapes [$p = 1$ circles, $p = 2$ squares, $p = 3$ diamonds, $p = 4$ triangles].

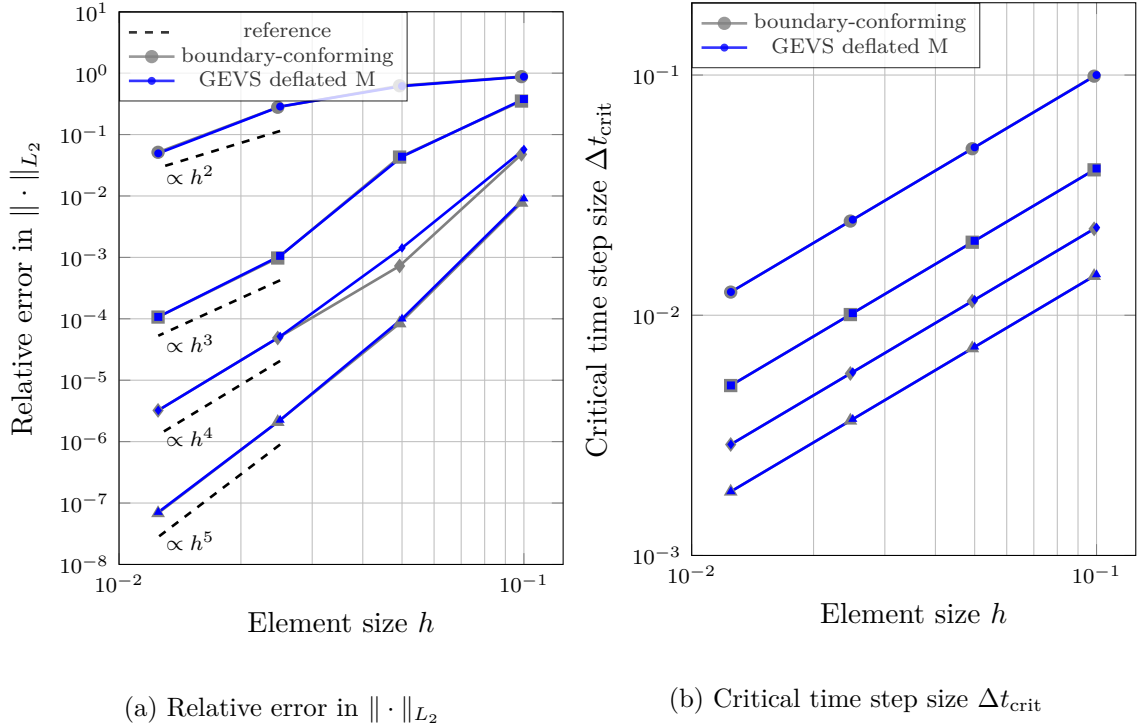

(a) Relative error in $\|\cdot\|_{L_2}$

(b) Critical time step size $\Delta t_{\text{crit}}$

Figure 7: Asymptotic accuracy and critical time step size of GEVS with Dirichlet boundary condition. Polynomial degrees $p = 1, \ldots, 4$ are represented by different marker shapes [$p = 1$ circles, $p = 2$ squares, $p = 3$ diamonds, $p = 4$ triangles].

**Critical time step size**

Finally, we investigate the behavior of the critical time step depending on the volume fraction in cut elements. For very small cuts, the mass matrix of the cut element tends to become singular, as the basis functions supported in the physical domain of the element become linearly dependent. To prevent GEVS from failing — as the corresponding eigenvalues of the generalized eigenvalue problem tend to infinity — we introduce a mild material stabilization ($\alpha = 10^{-10}$), even when employing the GEVS. We consider discretizations with 20 elements and varying polynomial degrees $p = 1, \ldots, 4$. The position of the immersed boundary $l_{\text{p}}$ is varied such that the volume fraction in the last element is logarithmically sampled between $10^{-8}$ and $10^{-1}$ using 15 sample points. Again, we impose an initial Gaussian distribution (53) and choose the simulation time $T = \frac{2l_{\text{p}}}{c}$ to obtain the initial distribution for a Neumann boundary condition and the flipped initial distribution for a Dirichlet boundary condition. For the time integration, we use CDM with $2\,000\,000$ time steps. Alongside GEVS, we also apply pure material stabilization to preserve stability for vanishing cut ratios and systematically compare the results obtained by both stabilization techniques. Figures 8 and 9 depict the relative error in the $L_2$-norm at the final time step and the critical time step size with respect to the cut fraction in the intersected element for Neumann and Dirichlet boundary conditions, respectively.

In the Neumann case, we observe that MS provides a lower bound for the critical time step size. However, for unfavorable cut configurations, the critical time step size still decreases significantly. An estimate for the minimum critical time step size in terms of a modified CFL condition tailored to FCM is provided in [40]. If Dirichlet boundary conditions are imposed with Nitsche's method, MS is not bounding the critical time step size anymore. As the cut fraction approaches zero, the critical time step size also tends to zero. On the other hand, applying GEVS preserves the accuracy for both Neumann and Dirichlet boundary conditions. Moreover, GEVS ensures that the critical time step size remains stable across all cut fractions. *In particular, the stability limit becomes completely independent of the volume ratio in the cut element.*

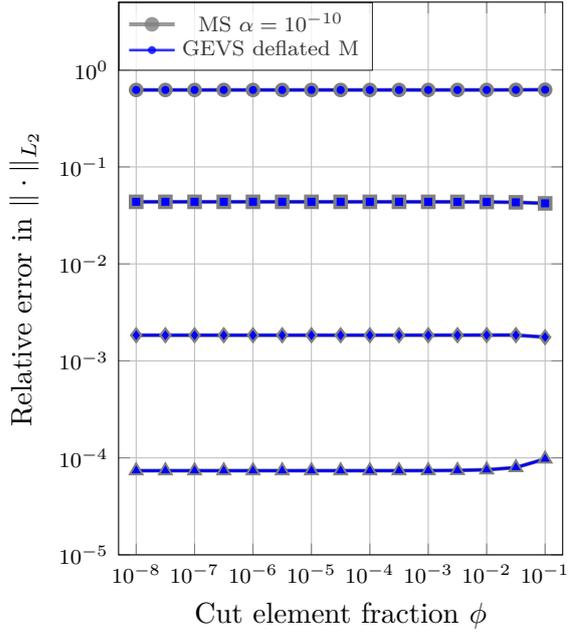 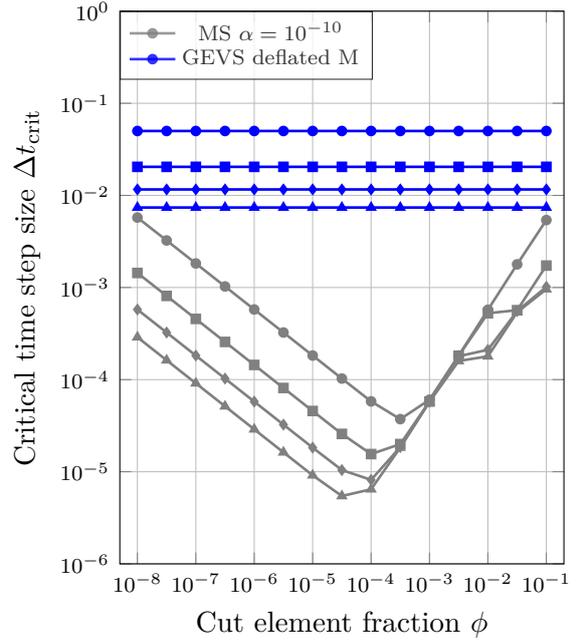

(a) Relative error in $\|\cdot\|_{L_2}$  
(b) Critical time step size $\Delta t_{\mathrm{crit}}$

Figure 8: Accuracy and critical time step size for varying cut size with Neumann boundary condition. Polynomial degrees $p = 1, \ldots, 4$ are represented by different marker shapes [$p = 1$ circles, $p = 2$ squares, $p = 3$ diamonds, $p = 4$ triangles].

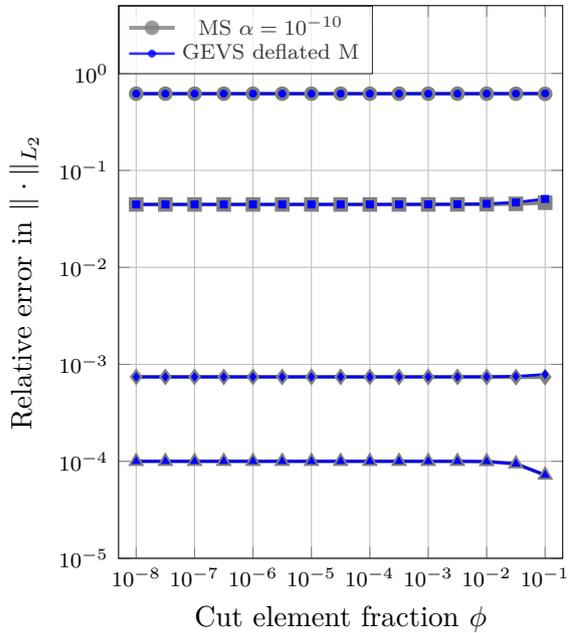 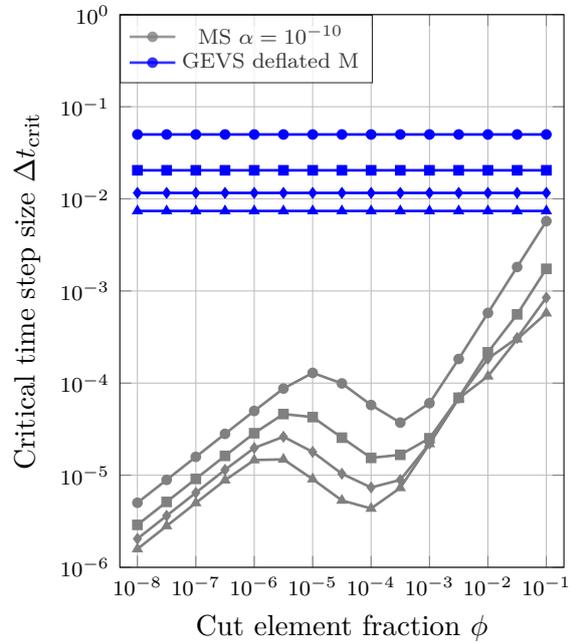

(a) Relative error in $\|\cdot\|_{L_2}$  
(b) Critical time step size $\Delta t_{\mathrm{crit}}$

Figure 9: Accuracy and critical time step size for varying cut size with Dirichlet boundary condition. Polynomial degrees $p = 1, \ldots, 4$ are represented by different marker shapes [$p = 1$ circles, $p = 2$ squares, $p = 3$ diamonds, $p = 4$ triangles].

# 4. Waves in an arc

**Problem statement**

In this section, we consider the two-dimensional immersed arc configuration illustrated in Figure 10. The setup parameters are shown on the right-hand side of the figure. The physical domain $\Omega$ is depicted in blue, while the fictitious domain $\Omega^\text{f}$ is shown in gray. The extended computational domain is $\Omega^\text{e} = [0, l] \times [0, l]$. Within the fictitious domain, the indicator functions takes the value $\alpha = 10^{-10}$. We investigate two cases. In the first case, homogeneous Neumann boundary conditions are imposed on the entire boundary $\partial\Omega$. In the second case, homogeneous Dirichlet boundary conditions are weakly enforced on the red parts of $\partial\Omega$ using Nitsche's method with a sufficiently high penalty value, uniform $\lambda_\text{E} = 10^7$. The material density is uniform $\rho = 1$, whereas the wave speed is graded linearly in radial direction ( $c = \frac{r}{r_\text{o}}$ with $r = \sqrt{(x-d)^2 + (y-d)^2}$ ) to ensure a straight/plane wave front. The initial condition is a circumferentially distributed Gaussian pulse:

$$\Psi_0(x,y) = 2\, e^{\frac{-(\theta-\theta_0)^2}{2\,\sigma^2}} \tag{53}$$

with $\theta = \arctan(\frac{y-d}{x-d})$, $\theta_0 = \frac{\pi}{4}$, and $\sigma = \frac{\pi}{40}$. Using the described setup, two waves propagate in opposite circumferential directions. At the final time $T = \frac{\pi}{2} - 2\,\theta_\Gamma$, the waves constructively interfere at the initial location. We obtain the analytical solution $\Psi(x,y,T) = \Psi_0(x,y)$ for the Neumann case, and $\Psi(x,y,T) = -\Psi_0(x,y)$ for the Dirichlet case. Figure 11 shows simulation results based on a reasonably fine discretization that illustrates this.

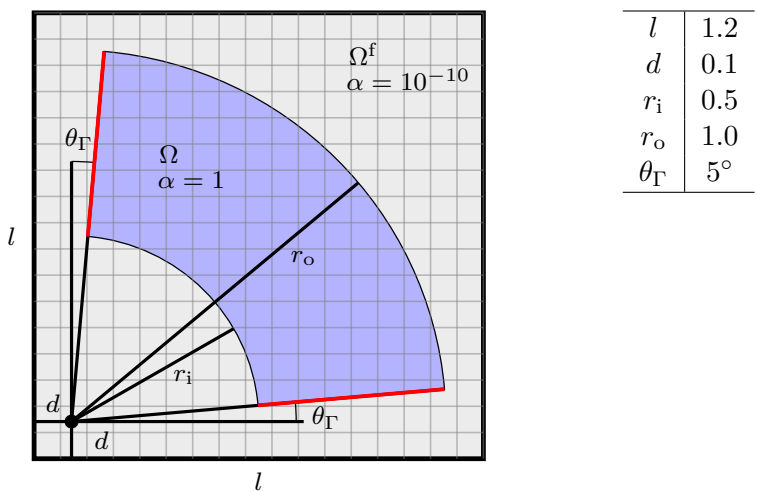

Figure 10: Computational domain and discretization for the test problem *waves in an arc* and configuration parameters.

**Asymptotic accuracy**

To assess the asymptotic accuracy, we perform an *h*-convergence study by logarithmically sampling the number of elements in each spatial direction, which discretize $\Omega^\text{e}$. We choose $n^\text{el} = [4, 8, 16, 32, 64, 128]$ for polynomial degrees $p = 1$ to $p = 4$. In the computational mesh, all elements that lie entirely within the fictitious domain $\Omega^\text{f}$ are discarded. For the integration of the system matrices and Nitsche terms, we employ the blended quadrature technique proposed by Kudela et al. [65]. Time integration is performed using 1 000 000 CDM steps. As in the previous section, we apply GEVS with mild material stabilization using $\alpha = 10^{-10}$. The desired eigenvalue $\lambda^*$ is defined as the largest eigenvalue among all full (uncut) elements. Figures 12 and 13 show the relative $L_2$-error at the final time step and the corresponding critical time step size for both the Neumann and Dirichlet cases. For reference, additional curves are included, depicting the pure material stabilization with $\alpha = 10^{-10}$ and expected convergence behavior $\propto h^{p+1}$.

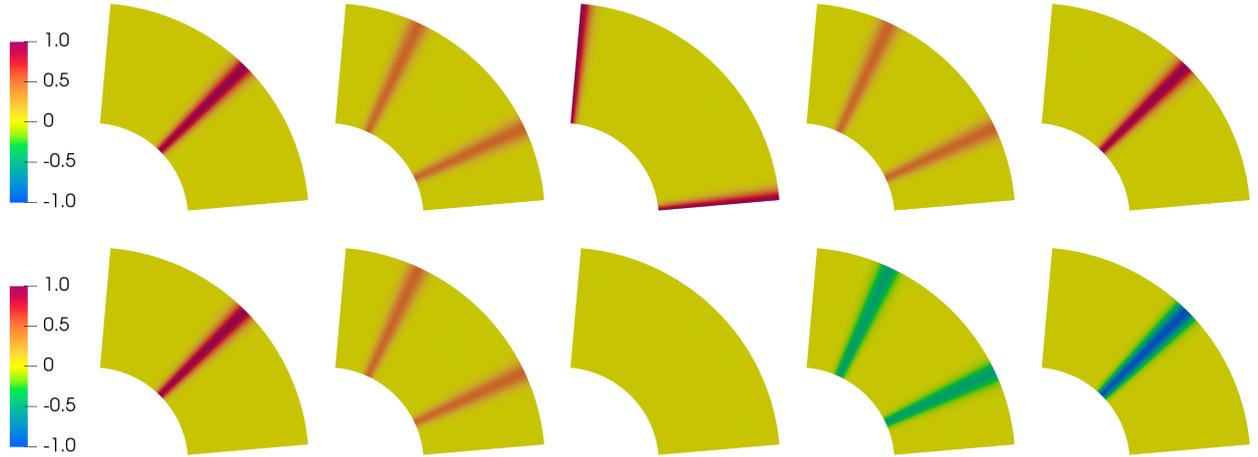

Figure 11: Snapshots of the simulations with Neumann boundary condition (top) and Dirichlet boundary condition (bottom) for $t = \frac{i}{4}T$, $i = 0, \ldots, 4$.

The results of the asymptotic study of the immersed arc are consistent with the one-dimensional example, meaning that GEVS efficiently eliminates outliers introduced by the immersed setup. Hence, optimal convergence is preserved, as the $L_2$-error scales with $\propto h^{p+1}$. In the Neumann case, the critical time step size increases by approximately one to two orders of magnitude, while in the Dirichlet case, the improvement reaches two to four orders of magnitude. Moreover, GEVS demonstrates strong robustness against large penalty factors in Nitsche's method (e.g., $\lambda_\mathrm{E} = 10^7$), which accurately enforce Dirichlet boundary conditions.

## 5. Conclusion

This paper presents a generalized eigenvalue stabilization (GEVS) strategy for immersed boundary finite element methods to eliminate the adverse effect on the stability caused by cut elements using explicit time integration. Formulated on element level, the approach relies solely on local information of the discretized system. In this work, we apply the strategy to spectral basis functions, i.e., $C^0$ continuous Lagrange polynomials, in combination with the material stabilization, which is typically utilized in the finite cell method. This particular combination of stabilization techniques guarantees definiteness of the generalized eigenvalue problem. However, the formulation is general in the sense that it can be readily employed in conjunction with other basis functions and stabilization techniques.

A global accuracy comparison demonstrates that GEVS is the only approach among the compared methods — (pure) material stabilization and regular eigenvalue stabilization of the mass matrix — that simultaneously preserves high accuracy in the lower part of the eigenvalue spectrum and fully eliminates spectral outliers. Further one-dimensional studies on asymptotic accuracy and the critical time step size for vanishing physical support in cut elements showed that GEVS retains optimal convergence rates, while restoring the critical time step sizes of equivalent uncut elements. These findings hold for both Neumann boundary conditions and Dirichlet boundary conditions weakly enforced with Nitsche's method or penalty formulations. Additional two-dimensional asymptotic studies on an immersed arc verify these findings. Therefore, we can conclude that GEVS preserves the accuracy of the finite cell method with mild (pure) material stabilization, while it may improve the critical time step size up to two orders in Neumann problems and up to four orders of magnitude in Dirichlet problems. Its robustness and high-accuracy make GEVS a viable stabilization strategy for efficient explicit dynamic simulations involving complex immersed geometries. However, the search for suitable mass-lumping schemes for cut elements remains unresolved—if, indeed, a method exists that can deliver high accuracy while yielding a diagonal mass matrix, is a question still under debate.

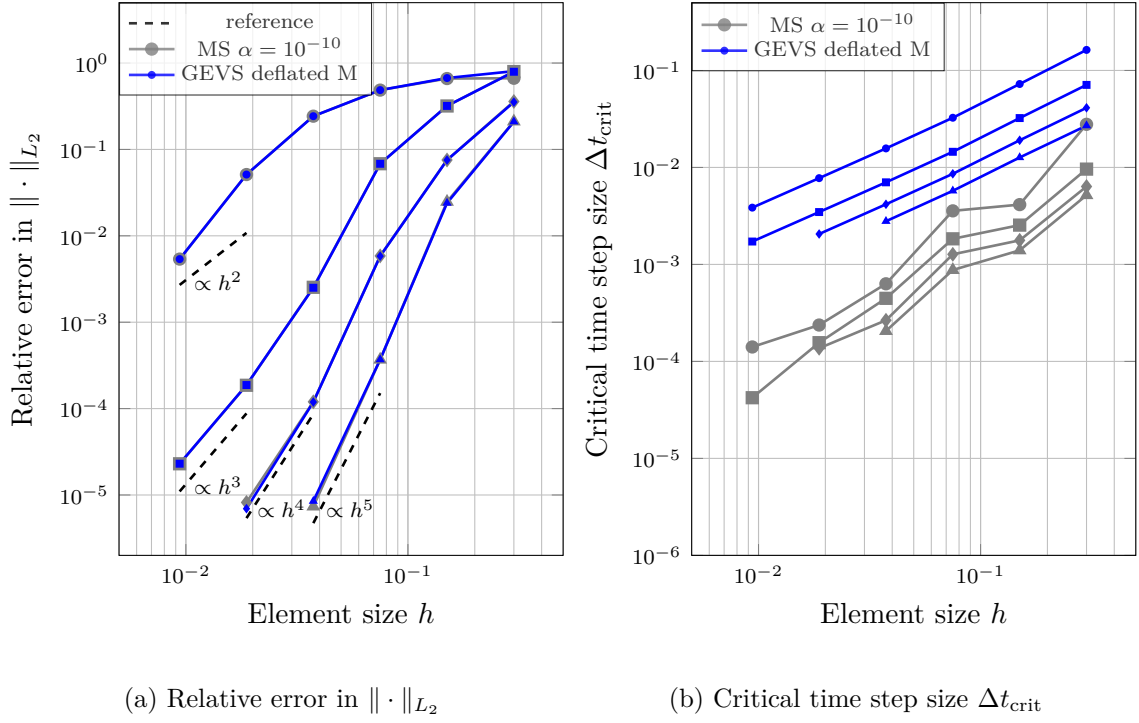

(a) Relative error in $\|\cdot\|_{L_2}$

(b) Critical time step size $\Delta t_{\text{crit}}$

Figure 12: Asymptotic accuracy and critical time step size wave in an arc with Neumann boundary condition. Polynomial degrees $p = 1, \ldots, 4$ are represented by different marker shapes [$p = 1$ circles, $p = 2$ squares, $p = 3$ diamonds, $p = 4$ triangles].

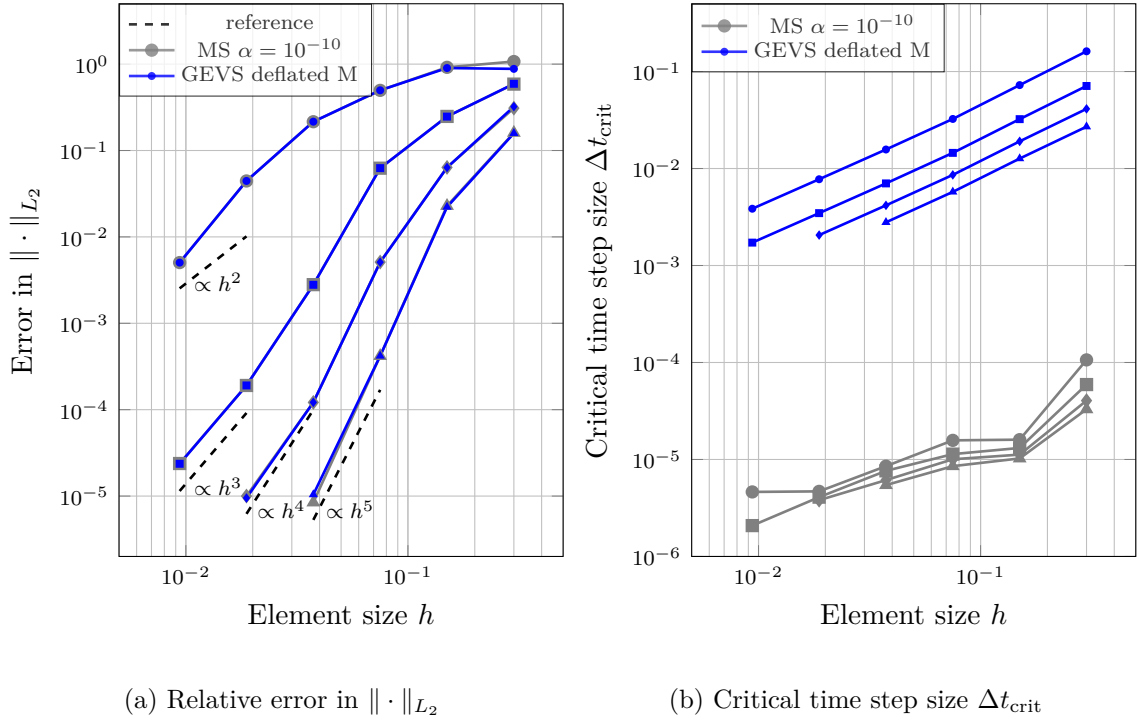

(a) Relative error in $\|\cdot\|_{L_2}$

(b) Critical time step size $\Delta t_{\text{crit}}$

Figure 13: Asymptotic accuracy and critical time step size wave in an arc with Dirichlet boundary condition. Polynomial degrees $p = 1, \ldots, 4$ are represented by different marker shapes [$p = 1$ circles, $p = 2$ squares, $p = 3$ diamonds, $p = 4$ triangles].

## A. Rank-one modification of the standard eigenvalue problem

Let $\mathbf{A} \in \mathbb{R}^{n \times n}$ be a real-valued symmetric matrix. The standard eigenvalue problem seeks eigenpairs $(\lambda_i, \boldsymbol{\phi}_i)$ such that

$$\mathbf{A}\boldsymbol{\phi}_i = \lambda_i \boldsymbol{\phi}_i \tag{54}$$

where $\lambda_i \in \mathbb{R}$ and $\boldsymbol{\phi}_i \in \mathbb{R}^n$ for $i = 1, \ldots, n$ denote the eigenvalue and corresponding eigenvector. Adding the outer product $\boldsymbol{\phi}_k \boldsymbol{\phi}_k^T$ to $\mathbf{A}$ results in an rank-one modification. The perturbed matrix is

$$\tilde{\mathbf{A}} = \mathbf{A} + \boldsymbol{\phi}_k \boldsymbol{\phi}_k^T. \tag{55}$$

Multiplying $\tilde{\mathbf{A}}$ with any eigenvector $\boldsymbol{\phi}_i$ yields

$$\tilde{\mathbf{A}}\boldsymbol{\phi}_i = (\mathbf{A} + \boldsymbol{\phi}_k \boldsymbol{\phi}_k^T)\boldsymbol{\phi}_i = \begin{cases} \mathbf{A}\boldsymbol{\phi}_i + \boldsymbol{\phi}_k \underbrace{\boldsymbol{\phi}_k^T \boldsymbol{\phi}_i}_{=0} = \underbrace{\lambda_i}_{=\tilde{\lambda}_i} \boldsymbol{\phi}_i & \text{if } i \neq k \\ \mathbf{A}\boldsymbol{\phi}_i + \boldsymbol{\phi}_k \underbrace{\boldsymbol{\phi}_k^T \boldsymbol{\phi}_i}_{=1} = \underbrace{(\lambda_i + 1)}_{=\tilde{\lambda}_i} \boldsymbol{\phi}_i & \text{if } i = k \end{cases}, \tag{56}$$

assuming the eigenvectors $\boldsymbol{\phi}_i$ are orthonormal, i.e., $\boldsymbol{\phi}_i^T \boldsymbol{\phi}_j = \delta_{ij}$. This shows that the rank-one modification, i.e., adding $\boldsymbol{\phi}_k \boldsymbol{\phi}_k^T$, shifts only $\lambda_k$ by one, while $\boldsymbol{\phi}_k$ and all other eigenvalues and eigenvectors remain unchanged. If we define

$$\tilde{\mathbf{A}} = \mathbf{A} + c_k \boldsymbol{\phi}_k \boldsymbol{\phi}_k^T \tag{57}$$

with

$$c_k = \lambda^* - \lambda_k, \tag{58}$$

we can shift the eigenvalue $\lambda_k$ corresponding to $\boldsymbol{\phi}_k$ to a desired value $\lambda^*$ without changing the rest of the spectrum. For a detailed discussion of deflation strategies considering the regular eigenvalue problem, see [52].

## B. Rank-one modification of the generalized eigenvalue problem

Let $\mathbf{A} \in \mathbb{R}^{n \times n}$ and $\mathbf{B} \in \mathbb{R}^{n \times n}$ be real-valued symmetric matrices. The generalized eigenvalue problem seeks eigenpairs $(\lambda_i, \boldsymbol{\phi}_i)$ satisfying

$$\mathbf{A}\boldsymbol{\phi}_i = \lambda_i \mathbf{B}\boldsymbol{\phi}_i, \tag{59}$$

where $\lambda_i \in \mathbb{R}$ and $\phi_i \in \mathbb{R}^n$ for $i = 1, ..., n$ denote the eigenvalues and corresponding eigenvectors. For the generalized eigenvalue problem, adding the term $\mathbf{B}\boldsymbol{\phi}_k \boldsymbol{\phi}_k^T \mathbf{B}$ to $\mathbf{A}$ leads to a rank-one modification. Multiplying the manipulated left-side matrix $\tilde{\mathbf{A}}$ with any eigenvector $\boldsymbol{\phi}_i$ yields

$$\tilde{\mathbf{A}}\boldsymbol{\phi}_i = (\mathbf{A} + \mathbf{B}\boldsymbol{\phi}_k \boldsymbol{\phi}_k^T \mathbf{B})\boldsymbol{\phi}_i = \begin{cases} \mathbf{A}\boldsymbol{\phi}_i + \mathbf{B}\boldsymbol{\phi}_k \underbrace{\boldsymbol{\phi}_k^T \mathbf{B}\boldsymbol{\phi}_i}_{=0} = \underbrace{\lambda_i}_{=\tilde{\lambda}_i} \mathbf{B}\boldsymbol{\phi}_i & \text{if } i \neq k \\ \mathbf{A}\boldsymbol{\phi}_i + \mathbf{B}\boldsymbol{\phi}_k \underbrace{\boldsymbol{\phi}_k^T \mathbf{B}\boldsymbol{\phi}_i}_{=1} = \underbrace{(\lambda_i + 1)}_{=\tilde{\lambda}_i} \mathbf{B}\boldsymbol{\phi}_i & \text{if } i = k \end{cases}, \tag{60}$$

assuming the eigenvectors $\boldsymbol{\phi}_i$ are $\mathbf{B}$-orthonormal, i.e., $\boldsymbol{\phi}_i^T \mathbf{B} \boldsymbol{\phi}_j = \delta_{ij}$. Thus, the rank-one modification shifts only the eigenvalue $\lambda_k$ by one while leaving $\boldsymbol{\phi}_k$ and all other eigenpairs unchanged. We can control the shift by choosing

$$\tilde{\mathbf{A}} = \mathbf{A} + c_k \mathbf{B}\boldsymbol{\phi}_k \boldsymbol{\phi}_k^T \mathbf{B} \tag{61}$$

with

$$c_k = \lambda^* - \lambda_k, \tag{62}$$

where $\lambda^*$ is a desired value without altering the rest of the spectrum.

Similarly, we can deflate the generalized eigenvalue problem by modifying the right-side matrix $\mathbf{B}$. Adding the rank-one modification $\mathbf{B}\boldsymbol{\phi}_k\boldsymbol{\phi}_k^T\mathbf{B}$ to $\mathbf{B}$ yields the perturbed matrix $\tilde{\mathbf{B}}$. The manipulated generalized eigenvalue problem becomes

$$\mathbf{A}\boldsymbol{\phi}_i = \tilde{\lambda}_i\tilde{\mathbf{B}}\boldsymbol{\phi}_i = \tilde{\lambda}_i(\mathbf{B} + \mathbf{B}\boldsymbol{\phi}_k\boldsymbol{\phi}_k^T\mathbf{B})\boldsymbol{\phi}_i = \begin{cases} \tilde{\lambda}_i\mathbf{B}\boldsymbol{\phi}_i + \tilde{\lambda}_i\mathbf{B}\boldsymbol{\phi}_k\underbrace{\boldsymbol{\phi}_k^T\mathbf{B}\boldsymbol{\phi}_i}_{=0} = \underbrace{\tilde{\lambda}_i}_{=\lambda_i}\mathbf{B}\boldsymbol{\phi}_i & \text{if } i \neq k \\ \tilde{\lambda}_i\mathbf{B}\boldsymbol{\phi}_i + \tilde{\lambda}_i\mathbf{B}\boldsymbol{\phi}_k\underbrace{\boldsymbol{\phi}_k^T\mathbf{B}\boldsymbol{\phi}_i}_{=1} = \underbrace{2\tilde{\lambda}_i}_{=\lambda_i}\mathbf{B}\boldsymbol{\phi}_i & \text{if } i = k \end{cases}, \quad (63)$$

assuming the eigenvectors are $\mathbf{B}$-orthonormal, i.e., $\boldsymbol{\phi}_i^T\mathbf{B}\boldsymbol{\phi}_j = \delta_{ij}$. This implies that $\tilde{\lambda}_k = \lambda_k/2$, while the remainder of the spectrum is not altered. If

$$\tilde{\mathbf{B}} = \mathbf{B} + c_k\mathbf{B}\boldsymbol{\phi}_k\boldsymbol{\phi}_k^T\mathbf{B} \quad (64)$$

and

$$c_k = \frac{\lambda_k}{\lambda^*} - 1, \quad (65)$$

the eigenvalue $\lambda_k$ corresponding to $\boldsymbol{\phi}_k$ takes the desired value $\lambda^*$.

In general, if we manipulate both matrices

$$\tilde{\mathbf{A}} = \mathbf{A} + c_{A,k}\mathbf{B}\boldsymbol{\phi}_k\boldsymbol{\phi}_k^T\mathbf{B} \quad \text{and} \quad \tilde{\mathbf{B}} = \mathbf{B} + c_{B,k}\mathbf{B}\boldsymbol{\phi}_k\boldsymbol{\phi}_k^T\mathbf{B}, \quad (66)$$

modified eigenvalue can be expressed as

$$\tilde{\lambda}_k = \frac{\lambda_k + c_{A,k}}{1 + c_{B,k}}. \quad (67)$$

For a detailed discussion of deflation strategies considering the generalized eigenvalue problem, see [52].

## Acknowledgements

We gratefully thank the Deutsche Forschungsgemeinschaft (DFG, German Research Foundation) for their support through the grants KO 4570/1-2 and RA 624/29-2 (both grant number 438252876), DU 405/20-1 (grant number 503865803), and EI 1188/3-1 (grant number 497531141).